\documentstyle[aps,pre,epsf,twocolumn,floats]{revtex}
\def\B.#1{{\bbox{#1}}}  
\def\C.#1{{\cal  #1}}
\begin{document}
\title{Statistically Preserved Structures in Shell Models
of Passive Scalar Advection}
\author{Yoram Cohen, Thomas Gilbert and Itamar Procaccia}
\address{Dept. of Chemical Physics,
The Weizmann Institute of Science, Rehovot 76100, Israel}
\maketitle
%%%%%%%%%%%%%%%%%%%%
\begin{abstract}
It was conjectured recently that Statiscally Preserved Structures
underlie the statistical physics of turbulent transport processes.
We analyze here in detail the time-dependent (non compact) linear
operator that governs the dynamics of correlation functions in
the case of shell models of passive scalar advection. The problem
is generic in the sense that the driving velocity field is neither
Gaussian nor $\delta$-correlated in time. We show how to naturally
discuss the dynamics in terms of
an effective compact operator that displays ``zero modes" which
determine the anomalous scaling of the correlation functions.
Since shell models have neither Lagrangian structure nor ``shape
dynamics" this example differs significantly from standard
passive scalar advection. Nevertheless with the necessary
modifications the generality and
efficacy of the concept of Statistically Preserved Structures
are further exemplified. In passing we point out a bonus of the 
present approach, in providing analytic predictions for
the time-dependent correlation functions in decaying turbulent
transport.
\end{abstract}
%%%%%%%%%%%%%%%%%%%%%% 
\section{Introduction}

Turbulent transport processes refer to the advection of a transported field
$\phi(\B.r,t)$ (scalar or vector) by a turbulent velocity
field $\B.u(\B.r,t)$ \cite{79MY,Csanady}. The basic equation of motion is linear,
having the form
\begin{equation}
\partial_t\phi={\cal L}\phi\, .
\label{basic}
\end{equation}
Here ${\cal L}$ is an operator that is built out of the
turbulent velocity field, and as such may be stochastic. Examples are the advection of a
passive scalar
$\theta(\B.r,t)$, with the equation of motion
\begin{equation}
\frac{\partial \theta}{\partial t} +\B.u\cdot \B.\nabla \theta=\kappa
\nabla^2\theta \ , \label{pas}
\end{equation}
or a vector, like a magnetic field $\B.B(\B.r,t)$ satisfying \cite{Zeldovich}
\begin{equation}
\frac{\partial \B.B}{\partial t} +(\B.u\cdot \B.\nabla) \B.B=(\B.B\cdot
\B.\nabla) \B.u + \kappa \nabla^2\B.B \ . \label{magnetic}
\end{equation}
We may also consider advection, as in \cite{01AP}, of a vector
field $\B.w$ whose divergence vanishes, $\nabla\cdot\B.w=0$:
\begin{equation}
\frac{\partial \B.w}{\partial t} +(\B.u\cdot \B.\nabla) \B.w=-\B.\nabla p
+ \kappa \nabla^2\B.w
\ .
\end{equation} 
In all these equations the velocity field $\B.u$ comes from either a solution
of a fluid-mechanical equation, or is a random field defined with some
statistical properties. A fundamental
consequence of the linearity of the equations of motion is that the
correlation functions may be expressed as
\begin{equation}
\begin{array}{l}
\langle \phi(\B.r_1,t) \ldots \phi(\B.r_N,t) \rangle = \vspace{5pt}\\ 
\hspace{40pt} \int \C.{\cal P}^{(N)}_{\underline{\B.r}|\underline{\B.\rho}}(t)\, \langle
\phi(\B.\rho_1,0) \ldots \phi(\B.\rho_N,0) \rangle\, d\underline{\rho}
\ ,
\end{array}
\label{propagator}
\end{equation} 
where $\langle\dots\rangle$ is an average over the statistics of 
the initial conditions {\em and} the statistics of the advecting
velocity field. The notation
$\underline{\B.r}=({\B.r}_1,\ldots , {\B.r}_N)$ is used for simplicity. Note that we have
used the passive nature of the transported field, i.e. the fact that the velocity is
independent of the {\em initial} distribution of $\phi$, to separate the averages over the
initial conditions and the velocity. Such a decoupling cannot be afforded 
at any other time because of the build-up of correlations between the
advecting and advected fields. The linear operator
${\cal P}^{(N)}_{\underline{\B.r}|\underline{\B.\rho}}(t)$ propagates the $N$th-order
correlation function from time zero to time
$t$. 

The evolution operator $\C.L$ generally includes dissipative terms, and without
fresh input (forcing) 
the statistics of the field $\phi$ is time-dependent; this is the {\em
decaying} case, Eq. (\ref{basic}). A related problem of much experimental
and theoretical interest is {\em forced} turbulent transport where an
input term $f$ is added to the Eq. (\ref{basic}). The situations
of interest in turbulence typically involve an input acting only at
large scales of order $L$. The objects of major interest are the
stationary correlation functions $F^{(N)}$ of the advected field,
\begin{equation}
F^{(N)}(\B.r_1,\dots,\B.r_N) \equiv \langle \phi({\B.r}_1,t)\cdots\phi({\B.r}_N,t)\rangle_f \
.
\label{defFN}
\end{equation}
One cares about the scaling properties
at distances much smaller than $L$ and at the
stationary state.  As usual in turbulent flows, the correlation functions of
the advected field are
expected to contain anomalous contributions behaving as
\begin{equation}
\label{dilation}
\langle \phi(\lambda{\B.r}_1,t)\cdots\phi(\lambda{\B.r}_N,t) \rangle_f
=\lambda^{\zeta_N} \langle \phi({\B.r}_1,t)\cdots\phi({\B.r}_N,t)
\rangle_f ,
\end{equation}
with scaling exponents $\zeta_N$ which cannot be inferred from
dimensional analysis. 

Recently \cite{01ABCPV}, two conjectures were proposed, pertaining to a wide variety
of turbulent transport processes, without special provisos on the
properties of the advecting velocity field:\\~\\
\noindent (i) In the decaying case, despite the non-stationarity of
the statistics, there exist special functions $Z^{(N)}(\underline{\B.r})$
such that
\begin{equation} 
I^{(N)}(t) = \int Z^{(N)}(\underline{\B.r})\, \langle \phi(\B.r_1,t) \ldots
\phi(\B.r_N,t) \rangle \, d\underline{\B.r} \;
\label{integrals}
\end{equation}
are statistical integrals of motion. In the limit of infinitely large
system $I^{(N)}$ does not change
with time.  It follows from
(\ref{propagator}) and the conservation of
$I^{(N)}(t)$ that in the infinite size limit the $Z^{(N)}$'s are left-eigenfunctions
of the operator:
\begin{equation}
Z^{(N)}(\underline{\B.r})=\int
\C.{\cal P}^{(N)}_{\underline{\B.\rho}|\underline{\B.r}}(t)
Z^{(N)}(\underline{\B.\rho})\,d\underline{\B.\rho} \ . \label{eig}
\end{equation}
note that this does not mean that the operator
$\C.{\cal P}^{(N)}_{\underline{\B.\rho}|\underline{\B.r}}(t)$ admits
an eigenvector decomposition, and see below for a further discussion of this
point.
\vskip .2cm
\noindent 
(ii) The anomalous part of the stationary correlation
functions in the forced problem is dominated by statistically
conserved structures.  In other words, at least in the
scaling sense
\begin{equation}
F^{(N)}(\underline{\B.r}) \sim Z^{(N)}(\underline{\B.r}) \ . \label{conj2}
\end{equation}
A direct
consequence is that the small-scale statistics of the transported field $\phi$ in the
forced case rests on the understanding of the decaying problem. A by-product is that the
scaling exponents $\zeta_N$ are universal, i.e.  independent of the
forcing mechanisms for any forcing that is statistically independent of
the velocity field. 

The conjectures were exemplified in the context of shell models of
passive scalar advection. The model's 
equations read \cite{01ABCPV,Luca}
\begin{eqnarray}
{d\theta_m\over dt}&=&i\big(k_{m+1}\theta_{m+1}u_{m+1}
+k_m\theta_{m-1}u^*_{m}\big)-\kappa k_m^2\theta_m ,
\label{passive}\\
&\equiv&{\cal L}_{m,m'} \theta_{m'} \nonumber
\end{eqnarray}
where the variables $u_n$ are
generated by the ``Sabra'' shell model \cite{Sabra}
\begin{eqnarray} \label{sabra}
\frac{d u_n}{dt}&=&i\big( ak_{n+1}  u_{n+2}u_{n+1}^*
 + bk_n u_{n+1}u_{n-1}^*  \\ \nonumber
&& +ck_{n-1} u_{n-1}u_{n-2}\big)  -\nu k_n^2  u_n +f_n\ .
\end{eqnarray}
Here the coefficients $a$, $b$, and $c$ are real. 
In Eqs. (\ref{passive}) and (\ref{sabra}) the wavevectors are $k_n=k_0 2^n$. 
The velocity forcing $f_n$ is limited to the first  shell $n=0$.
For $\kappa=\nu=0$ and $a+b+c=0$ the energies $\sum_n |u_n|^2$ and $\sum_n |\theta_n|^2$ are
{\em dynamically conserved}, i.e. realization by realization. The statistical
physics of this model were studied carefully \cite{Sabra} in the regime of $b\approx -0.5$.
Taking the forcing to be random (with random phases) leads to non-trivial
statistics of the velocity field, with anomalous exponents that
characterize the scaling behavior of the correlation functions.

The operator $\B.{\cal P}^{(N)}$
of Eqs. (\ref{propagator}), (\ref{eig}) takes here the explicit form
\begin{equation} 
\C.{\cal P}^{(N)}_{\underline{n}|\underline{m}}(t)=\langle R_{n_1,m_1}(t|0)
\cdots R_{n_N,m_N}(t|0)\rangle \ ,
\end{equation}
where $\underline {n}= (n_1,\dots, n_N)$ and
\begin{equation}
R_{n,m}(t|0)\equiv T^+\left[\exp[\int_{0}^{t}ds {\cal L}(s)]\right]_{n,m}\ , \label{Rnm}
\end{equation}
with $T^+$ being the time ordering operator. Note that for notational
simplicity we dropped the dependence on the initial time from $\B.{\cal P}^{(N)}$,
but left it, for future purposes, in $\B.R(t|0)$.

To demonstrate the {\em statistical} conservation laws, two
things were done \cite{01ABCPV}. First the forced problem was considered, adding 
random forcing to Eq. (\ref{passive}):
\begin{eqnarray}
{d\theta_m\over dt}&=&{\cal L}_{m,m'} \theta_{m'}+f_m\ , \label{withf}\\
\langle f_m(t) f^*_n(t')\rangle &=& C_m\delta_{m,n}\delta(t-t') \ . \label{ff}
\end{eqnarray}
 Due to phase symmetry
constraints \cite{Sabra}, there is only one non-zero second order correlation, but
a number of different higher order ones. For example, the 
correlation $\langle \theta_{n+2}\theta^*_{n+1}\theta^*_{n+1}\theta_{n-1}\rangle_f$ is not
zero.  For concreteness
we concentrated our attention to the following ones (we put a subscript
$f$ to stress that these are statistical averages in the stationary forced ensemble):
\begin{eqnarray}
F^{(2)}_n&\equiv& \langle |\theta_n|^2\rangle_f\ , \label{F2def}\\
F^{(4)}_{n,m}&\equiv&  \langle |\theta_n|^2|\theta_m|^2\rangle_f\label{F4def} \ , \\
%F^{(4,2)}_n&\equiv&  \langle \theta_{n+2} \theta_{n+1}^* \theta_{n+1}^*\theta_{n-1}\rangle_f\ , \\ 
F^{(6)}_{n,m,k}&\equiv&  \langle |\theta_n|^2|\theta_m|^2
|\theta_k|^2\rangle_f \ .
\end{eqnarray}

Secondly, the decaying problem was examined,
 preparing initial states $\theta_n(t=0)$
and following their evolution. 
The following objects were then computed:
\begin{eqnarray}
I^{(2)}(t)&\equiv& \sum_n \langle |\theta_n(t)|^2\rangle ~F^{(2)}_n\label{I2def} \ ,\\
I^{(4)}(t)&\equiv& \sum_{n,m} \langle |\theta_n(t)|^2|\theta_m(t)|^2\rangle 
~F^{(4)}_{n,m} \label{I4def} \  ,\\
I^{(6)}(t)&\equiv& \sum_{n,m,k} \langle 
|\theta_n|^2|\theta_m(t)|^2|\theta_k(t)|^2\rangle ~F^{(6)}_{n,m,k} \ .
\end{eqnarray}
%%%%%%%%%%%%%%%%%%%%%%%%%%%%%%
\begin{figure}
\epsfxsize=7 truecm
\epsfbox{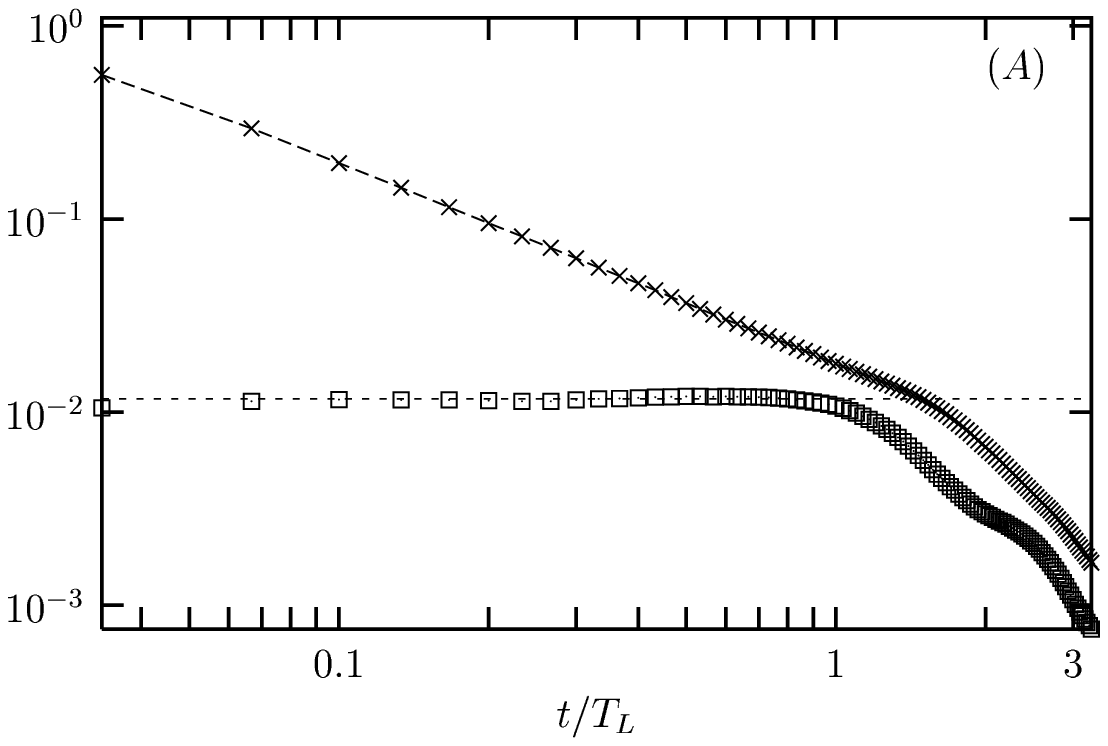}
\epsfxsize=7  truecm
\epsfbox{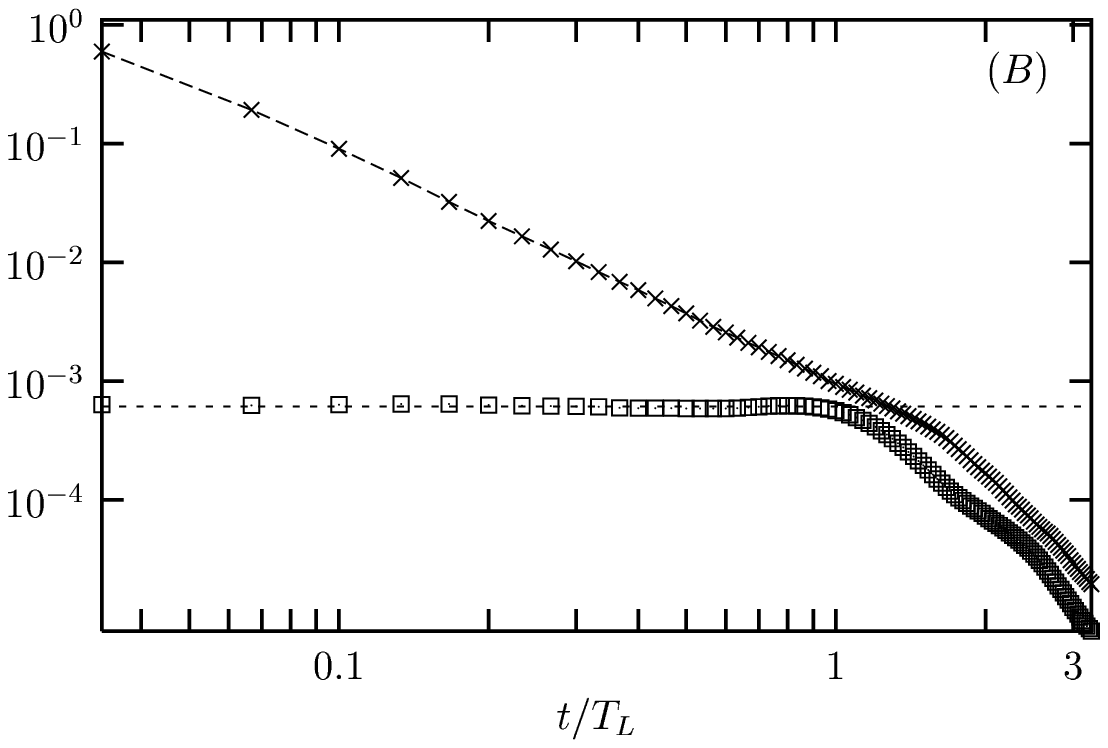}
\epsfxsize=7 truecm
\epsfbox{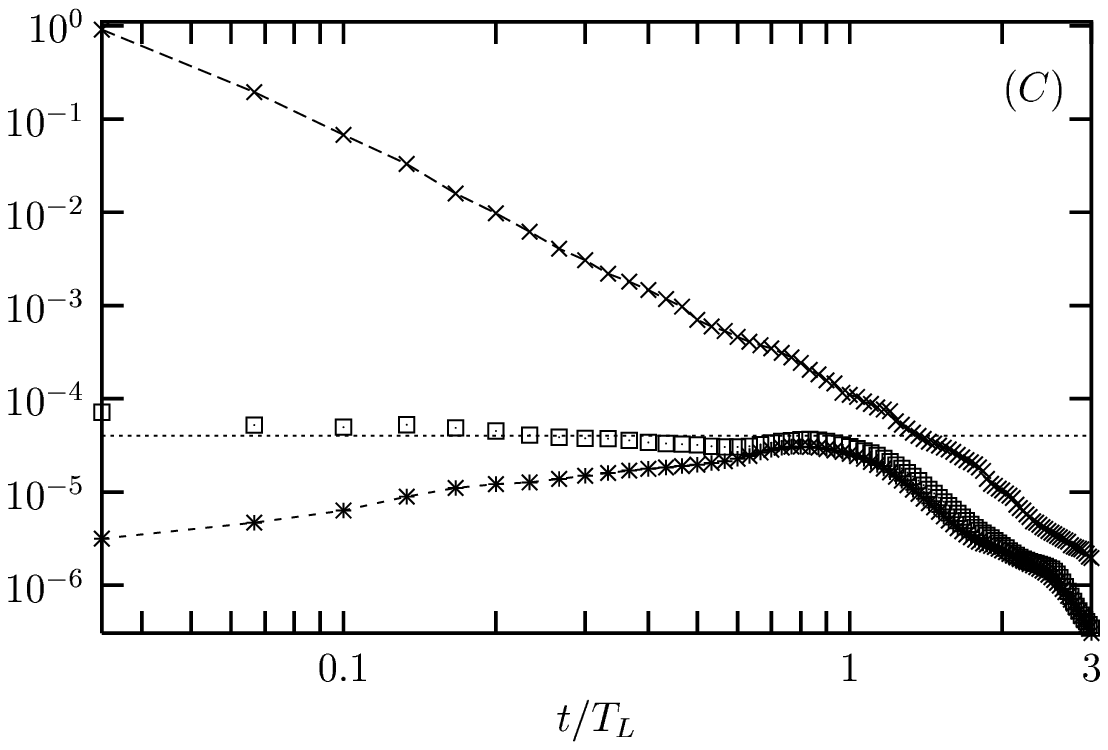}
\caption{Panel (A): time dependence of the decaying second  order
correlation functions ($\times$), together with the time dependence of 
the statistically conserved
quantities $I^{(2)}$, ($\Box$). Equations 
(\ref{passive}) and  (\ref{sabra}) have been integrated with a total
number of shells $N=33$. Time in the horizontal-axis is given in units
of the large eddy turn over time $T_L$. Panel (B): the same as panel (A)
but for the fourth order correlation function and with $N=25$. 
Panel (C): the same of
panel (B) but for the sixth order correlation function. Here we also present
$I^{(6)}$  when we replace the forced solution
$F^{(6)}_{n,m,k}$ with its dimensional prediction, ($\ast$) .
In the simulations $\kappa=\nu=5\times 10^{-7}$, $a=1$, $b=-0.4$, and
$c=a+b$.   The wavevectors are $k_n=k_0\,2^m$ with $n=0,\dots,N$. The smallest wavevector
is given by $k_0=0.05$ while $N$ defines the ultraviolet cut-off.
As initial states distributions of
$\theta_n=0$ were taken, except for $n= 14,15$ where the 
field was initialized with a constant modulus and random phases. The random
forcing of the passive scalar was restricted to the first shell.}
\end{figure}
%%%%%%%%%%%%%%%%%%%%%%%%%%%%%%%%%%%%%
Figure 1 summarizes the results which are reprodcued from \cite{01ABCPV}. We show, for these
three orders, (i) the time dependence of the $n$th order decaying correlation functions
themselves, (ii) the time dependence of $I^{(N)}(t)$. In panel (C) we show also for
comparison the time dependence of
$I^{(6)}(t)$ if we replace the measured forced $F^{(6)}$ by its dimensional shell dependence
(i.e. the shell dependence if the Kolmogorov theory were right). We see that only the
properly computed
$I^{(n)}(t)$ are time independent for times smaller than the large
scale eddy turn over time,  $T_L$.
The decay observed for  times larger than $T_L$ is simply due to finite
size effects intervening when the decaying field reaches the largest
scales. 

In trying to understand these results, it is very tempting to interpret
Eq. (\ref{eig}) as an eigenvalue equation, with $Z^{(N)}$ being an
eigenfunction of eigenvalue 1. Unfortunately, the operator 
$\C.{\cal P}^{(N)}$ is not Hermitian, and in addition it
does not lend itself to an expansion in terms of eigenvectors and
eigenvalues: it is not defined on a compact space. There are two
``non-compact" directions, that of length scale and that of time.
We thus need to learn how to take care of these before we can
write down a proper theory.

In the context of the passive scalar advection problem, Eq. (\ref{pas}), 
these issues were solved elegantly in the framework of Lagrangian
dynamics \cite{GZ,98GPZ,98FMV,98BGK,00AP}. For the passive scalar 
equation (\ref{pas}) the
advected field is conserved along the trajectories of the tracer particles
$d{\B.r}(t)=\B.u({\B.r}(t),t)\,dt+\sqrt{2\kappa}\,d\B.\beta(t)$, where
$\B.\beta(t)$ is a Brownian process. To know the scalar
field at position $\B.r$ and time $t$ it is enough to track the
corresponding tracer particle back to its initial position $\B.\rho$.  The
evolution operator $\C.{\cal P}^{(N)}_{\underline{\B.r}|\underline{\B.\rho}}(t)$ in
Eq. (\ref{propagator}) coincides then with the probability density that
$N$ tracer particles reach the positions $\underline{\B.r}$ at time
$t$ given their initial positions $\underline{\B.\rho}$. For example,
to understand the exponent $\zeta_3$ one needs to focus on the
dynamics of three tracer particles.  Obviously, three particles define
at any moment of time a triangle, which in turn is fully
characterized by one length scale $R$ (say the sum of the lengths of
its sides), two of its internal angles, and all the angles that
specify the orientation of the triangle in space.  When the particles
are advected by the turbulent velocity field, the scale $R$ of the
triangle and its shape (angles) change continuously. The statement
that can be made is that {\em there exist distributions on the space
of the triangle configurations, that are statistically invariant to
the turbulent dynamics} \cite{GZ,98GPZ,98FMV,00AP}. In other words, 
if we release trios of
Lagrangian tracers many times into the turbulent fluid, and we choose
the distribution of their shapes and sizes correctly, it will remain
invariant to the turbulent advection\cite{00CV}. Such statistically conserved
structures are the aforementioned zero modes and they come to dominate
the statistics of the scalar field at small scales. The anomalous
exponents of the zero modes, such as $\zeta_3$, can be understood as
the rescaling exponents characterizing precisely such special
distributions. Of course, the same ideas apply to any order
correlation function with the appropriate shape dynamics. The
relevance of Lagrangian trajectories can be also demonstrated for
the magnetic field case (\ref{magnetic}), by adding a tangent vector
to the tracer particle, and see \cite{CM} for more details.

The problem of non-compactness due to the explicit time dependence of the
operator is taken
care here by expressing time in terms of a single scale variable $R$, using the
Richardson law of turbulent diffusion \cite{98BGK}. Then instead of looking at the problem on
the non-compact space of particle separation, one focuses on the space of shapes
which is compact, and in which one can demonstrate the existence of
eigenfucntions and eigenvalues \cite{98BGK,00AP}. Obviously, for the case of the shell
model considered here we cannot repeat verbatim the same ideology. There are no
``shapes", and it is not immediately obvious how to relate time to scales.
The Lagrangian invariance is broken by the discretization of shell space,
and the genericity of the time properties of the velocity field does
not allow explicit calculations of the operator
$\C.{\cal P}^{(N)}_{\underline{\B.m}|\underline{\B.n}}(t)$.
 
The aim of this paper is to achieve the equivalent understanding for
the shell model, that in \cite{01ABCPV} was orginally chosen to be as
far removed as possible from the continuous passive scalar problem. We 
will discover that also in this case there is a typical
``moving" scale that carries the explicit time dependence. By considering
the relevant operators with shell indices expressed 
in terms of the moving
scale, we compactify the picture with respect to its time dependence. Moreover, in this
moving frame we will discover that the operators decay rapidly
as a function of shell differences.  This will allow us to compactify the
theory altogether and to offer a satisfactory understanding
of the existence of the statistically preserved structures and its
implication for the forced problem.

In Sect. 2 we present the theory for 2nd order objects. On the basis
of numerical simulations we offer an analytic form for the 
operator $\B.{\cal P}^{(2)}$. We show that it has an explicit
time dependence in addition to a dependence on a moving scale
that we identify analytically. In Sect. 3 we use the explicit
form of $\B.{\cal P}^{(2)}$ to explain why $I^{(2)}$ is a statistical
constant of the motion. The basic property that is crucial is the
effective compactness of the operator in the space of shells, once
it is expressed in terms of the moving scale. Next we show how the
forced stationary correlation function $F^{(2)}$ is obtained
by solving the forced problem with the same propagator $\B.{\cal P}^{(2)}$.
Finally we derive the fact that $F^{(2)}$ acts as a
left-eigenvector of $\B.{\cal P}^{(2)}$ with eigenvalue 1.
To help in throwing light on some issues we also consider in this 
section a simple model obtained by replacing the Sabra model
for the velocity field by a delta-function
correlated field (the Kraichnan shell model \cite{Luca,RHK}). 
In Sects. 4 and 5 we turn to a discussion of the 4th order objects.
We proceed in parallel to what had been achieved in Sects. 2
and 3. We first derive, on the basis of simulations and the 
fusion rules \cite{96LP}, the analytic form of $\B.{\cal P}^{(4)}$. Using
this form we explain why $I^{(4)}$ is a statistical constant of the
motion when the stationary correlation function $F^{(4)}$ is 
identified with $Z^{(4)}$. Last we turn to the forced problem,
and demonstrate that $F^{(4)}$ is indeed the forced solution.
This calculation is not trivial, calling for a careful discussion
of the time-decay and decorrelation properties of the operators 
$R_{n,m}(t|0)$. Throughout the discussion we make use of the
simpler Kraichnan shell model in which the operators are
all computed analytically (see Appendix) to further our understanding of 
the generic case. In Sect. 6 we present a discussion and 
a summary of the paper. One very important conclusion is that
we can in fact offer an {\em analytic solution} for the time-
dependent correlation functions in the decaying case; this
is a considerable bonus of the present approach.
%%%%%%%%%%%%%%%%%%%%%%%%%%%%%%%%%%%%%%%%%%%%%%%%%%
\section{The form of the 2nd order time propagator}
\subsection{Simulations}
In this section we analyze the form of the 2nd order propagator
that governs the dynamics of the second order passive structure
function. It is defined by:
\begin{equation}\label{Pdef}
\langle |\theta_n(t)|^2\rangle=\sum_m {\cal P}^{(2)}_{n|m}(t)\langle
|\theta_m(0)|^2\rangle\, .
\end{equation}
The $\langle \dots \rangle$ average is over realizations of the
velocity field and the initial conditions of the passive field.
As mentioned above at time $t=0$ the statistics of the advected
field is independent of the statistics of the velocity field. 
Using simulations we can generate the matrix representation of
${\cal P}^{(2)}_{n|m}(t)$ column by column by initiating a decaying
simulation (without forcing) starting with $\delta$-function
initial conditions in shell $m$. Measuring $\langle |\theta_n(t)|^2\rangle$
and averaging over many realizations of the Sabra velocity field
we collect data for ${\cal P}^{(2)}_{n|m}(t)$.

\begin{figure}
\hskip -0.5cm
\epsfxsize=8truecm
\epsfysize=8truecm
\epsfbox{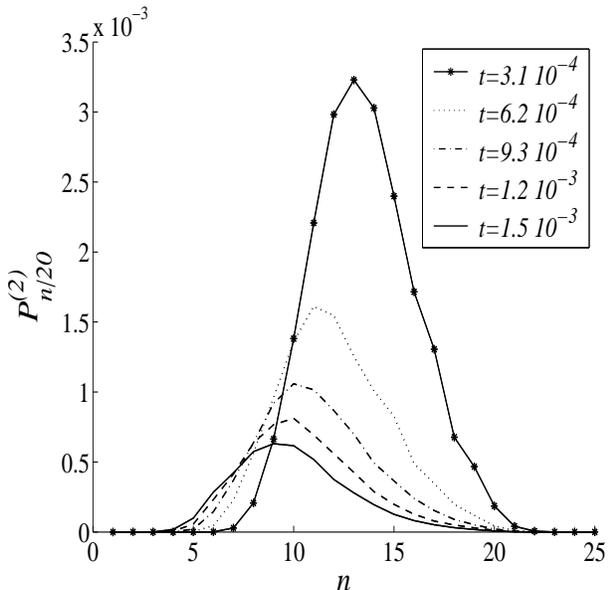}
\caption{Typical time dependence of one column of the second 
order propagator ${\cal P}^{(2)}_{n|m}(t)$. Shown here is ${\cal P}^{(2)}_{n|20}(t)$
for the different times displayed in the inset in units of $\tau_0$.
Note that the maximum moves in time to lower shell numbers.}
\label{Pnm}
\end{figure}

In Fig. \ref{Pnm} we show a typical column of ${\cal P}^{(2)}_{n|m}(t)$, where
$m=20$. We used 28
shells in both Sabra and the passive field, with the dissipative
scales being around $n=25$.
%%%%%%%%%%%%%%%%%%%%%%%%%%%%%
%%%%%%%%%%%%%%%%%%%%%%%%%
We observe two effects. First, the overall area under the curve decreases with
time, this is the effect of the dissipation. Second, the maximum in the
curve shifts to lower shell numbers. These are the two issues that we
need to tackle, the time dependence and the increase in length scale
(or, equivalently the decrease in shell number),
which contribute to the non-compact nature of our operator.

In attempting to contain these two issues we try the following
ansatz for the propagator:
\begin{equation}\label{Pform}
{\cal P}^{(2)}_{n|m}(t)=\frac{\tau_m}{t}H(n-\tilde m(t,m))\ , \quad {\rm for~ } t\gg\tau_{m} \ ,
\end{equation}
where $\tau_{m}$ is a typical time scale associated with the shell in which
the simulation was initiated. We use below
\begin{equation}
\tau_{m} = 2^{-m \zeta_2}/k_0 \sqrt{\langle |u_0|^2\rangle}  \ , \label{taun}
\end{equation}
with $\zeta_2$ being the scaling exponent of the second order structure function,
cf. Eq. (\ref{dilation}). 
Accordingly, all times $t$ below are also
measured in units of $\tau_0=1/ k_0 \sqrt{\langle |u_0|^2\rangle}$.
 The function $H(x)$ has a peak at $x=0$, with $ H(0)=1$ and for $x>0$ it has the
form 
\begin{equation} \label{Fscale}
H(x)\sim 2^{-\zeta_2 x}\, ,
\end{equation} 
The location of the maximum  of ${\cal P}^{(2)}_{n|m}(t)$ is $\tilde m(t,m)$, and is a
real valued function of time and of the initial peak location for $t=0$, which is $m$. 
For $t>0$ it satisfies  $\tilde m(t,m) < m$.

\begin{figure}
\hskip -0.5cm
\epsfxsize=8truecm
\epsfysize=8truecm
\epsfbox{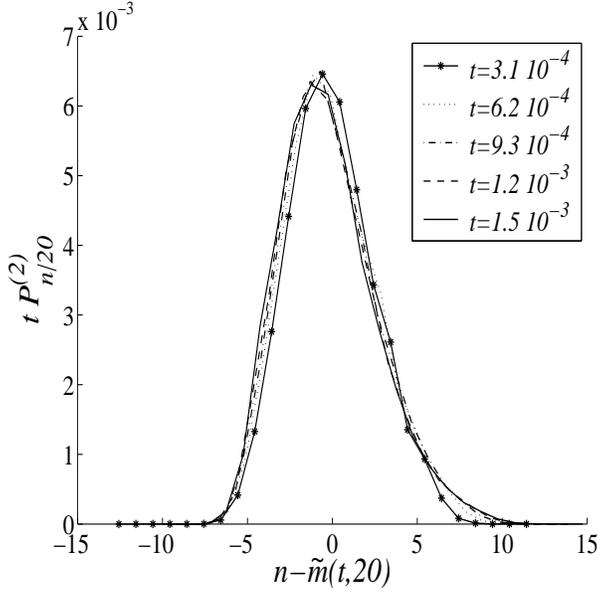}
\caption{A plot of $t {\cal P}^{(2)}_{n|20}(t)$ as a function of $n-\tilde m(t,20)$. 
The quality of the data collapse is deteriorating at the right tail 
because of viscous effects, where power law scaling crosses over to exponential decay.}
\label{Pnmt}
\end{figure}

To show that the ansatz (\ref{Pform}) is well supported by the data, 
we show in Fig. \ref{Pnmt} $t {\cal P}^{(2)}_{n|m}(t)$ as a function of $n-\tilde m(t,m)$.
%%%%%%%%%%%%%%%%%%%%%%%%%%%%
%%%%%%%%%%%%%%%%%%%%%%%%% 
The quality of the data collapse speaks for itself. We draw the attention
of the reader to the fact that the function shown in Fig. \ref{Pnmt} falls off sharply 
around the maximum. This will be the clue to understanding how to remove
the noncompact dependence on the ever increasing scale $\tilde m(t, m)$.
Sums over $n$ will be extended below from $-\infty$ to $\infty$, with impunity.
The important conclusion is that to a good approximation the following sum
\begin{equation}
\label{sumofF}
\sum_{n} H(n-\tilde m)=\sum_{n-\tilde m}  H(n-\tilde m) =\sum_k H(k)
={\rm Const} \ , \label{Hconst}
\end{equation}
is time independent. 
\subsection{The time dependence of the maximum}
Next we want to find an analytic expression for the moving scale $\tilde m(t,m)$.
In order to find the time behavior of the peak we examine Eq. (\ref{Pdef})
for an initial condition $\langle |\theta_k(0)|^2\rangle=\delta_{k,m}$. On the
one hand, for
these initial conditions, after time differentiating we get: 
\begin{equation}\label{disp1}
\frac{d}{dt}\sum_n\langle |\theta_n(t)|^2\rangle=\frac{d}{dt} \sum_n
{\cal P}^{(2)}_{n|m}(t)\, .
\end{equation}
On the other hand, using Eq. (\ref{passive}) one finds:
\begin{equation}\label{disp2}
\frac{d}{dt}\sum_n\langle |\theta_n(t)|^2\rangle=-2 \kappa\sum_n
k_n^2\langle |\theta_n(t)|^2\rangle\, .
\end{equation}
To evaluate the sum on the R.H.S of Eq. (\ref{disp2}), we note that for this
linear problem the shell $d$ from which the dissipation of the scalar
becomes significant is independent of the scalar value (and thus time
independent). We can estimate it by comparing the terms on the RHS of
Eq. (\ref{passive}):
\begin{equation}
\kappa k^2_d \approx u_d k_d \ . \label{defkd}
\end{equation}
We can now estimate the scalar dissipation, under the approximation that it takes
place mainly in shells with $m>d$. In this region the value
of $\langle |\theta_n(t)|^2\rangle$ begins to fall off exponentially 
with $k_n$, 
and the sum in Eq. (\ref{disp2}) is well approximated by the first term 
$\kappa k^2_d \langle|\theta_d(t)|^2\rangle$. 
Plugging in the functional form of $\B.{\cal P}^{(2)}$ given by 
Eq. (\ref{Pform}), using Eqs. (\ref{sumofF})-(\ref{disp2}) we get:
\begin{eqnarray}
\frac{d}{dt}\sum_n {\cal P}^{(2)}_{n|m}(t)&=&-\frac{1}{t^2}\sum_k
H(k)\nonumber\\&\approx&-\frac{c
  \kappa k_d^2 \tau_m}{t}2^{-\zeta_2(d-\tilde m(t,m))} \ . \label{setm}
\end{eqnarray}
Examining Eq. (\ref{setm}) we conclude that in order for the RHS to
scale like $t^{-2}$ for $t\gg \tau_m$, while demanding that for $t\approx \tau_m$ $\tilde
m(t,m)\approx m$, we must have: 
\begin{eqnarray}\label{mform}
\tilde m(t,m)&=&m-\frac{1}{\zeta_2}\log_2\left[g\left(\frac{t}{\tau_{m}}\right)\right]\ , 
\nonumber\\
g(0)&=&1\ , \quad \lim_{x\to\infty} g(x) =x \ .
\end{eqnarray}
where $\tau_{m}$ was defined in 
Eq. (\ref{taun}). Thus for large times we will use
\begin{equation}
\tilde m(t,m)=-\frac{1}{\zeta_2}\log_2(\frac{t}{\tau_{0}})\ , \quad
t\gg\tau_{m} \ ,
\end{equation}
 Note that for large time ($t\gg \tau_m$), $\tilde m(t,m)$ 
becomes
independent of $m$. This is approporiate, since the exponential
increase in typical time scales $\tau_m$ when the shell index decreases implies
that the position of the
maximum  becomes independent of its initial position.  We can now express
the time dependence of the operator  $\B.{\cal P}^{(2)}$ in Eq. (\ref{Pform}), solely through
the time behavior of $\tilde m(t,m)$, by inverting Eq. (\ref{mform}), to find $t$:
\begin{equation}\label{Pmform}
{\cal P}^{(2)}_{n|m}(t)\propto 2^{-\zeta_2(m-\tilde m)}H(n-\tilde m) \ ,
\!\!\quad t \ge\tau_{m} \ .
\end{equation}
Having done so, we have gotten rid altogether of the explicit
time dependence of ${\cal P}^{(2)}_{n|m}(t)$. Note that the depndence
of the operator on both its shell indices turns naturally to a
dependence on the difference between these indices and the single
moving shellThis is the first
important step in overcoming the non-compactness of our 2nd order operator.
%%%%%%%%%%%%%%%%%%%%%%%%%%%%%%%%%%%%%%%%%%
\section{Consequences of the form of the 2nd order propagator}

At this point we can reap the benefit of the explicit form
of the 2nd order propagator Eq. (\ref{Pmform}). First we derive
the existence of the statistical constant of the motion $I^{(2)}$.

\subsection{Second order constant of the motion}

\begin{figure}
\hskip -0.5cm
\epsfxsize=8truecm
\epsfysize=8truecm
\epsfbox{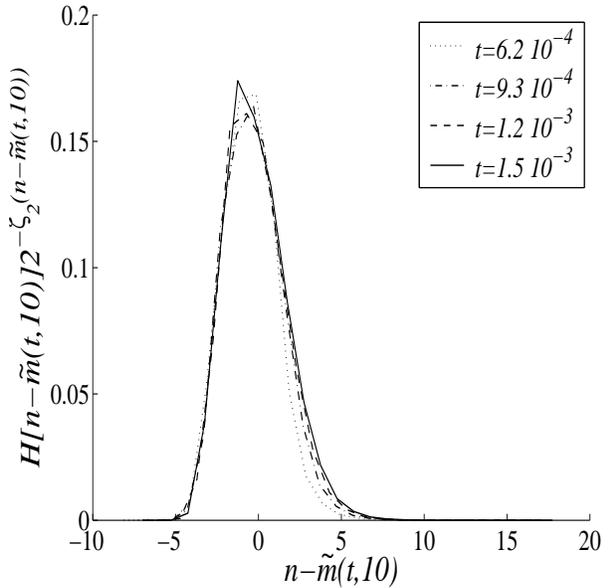}
\caption{The summand $H(n-\tilde m)2^{-\zeta_2 (n-\tilde m)}$ as a function of $n-\tilde m$.}
\label{summand}
\end{figure}

Returning to the definition of $I^{(2)}$, Eq. (\ref{I2def}), and recognizing that
$F^{(2)}_n\propto 2^{-\zeta_2 n}$ (which is also demonstrated in
the next Subsection), we see that we need to
evaluate the weighted sum
$\sum_n\langle |\theta_n(t)|^2\rangle 2^{-\zeta_2 n}$. Since the
problem is linear, any initial condition can be represented as a weighted
sum of $\delta$-function initial conditions, and therefore we only
need to consider sums of the form
\begin{equation}\label{evalsum}
\sum_n {\cal P}^{(2)}_{n|m}(t) 2^{-\zeta_2 n}=2^{-\zeta_2 m}\sum_n
H(n-\tilde m)2^{-\zeta_2 (n-\tilde m)}\, .
\end{equation}
As the components of the sum are a function of $n$ only through the
combination $n-\tilde m$, we can change the summation to run
on $n-\tilde m$. In light of Eq. (\ref{Hconst}) the sum is time independent. In
Fig. \ref{summand} we show the summand as a function of time and $n-\tilde m$.
%%%%%%%%%%%%%%%%%%%%
%%%%%%%%%%%%%%%%%%%
\subsection{The Forced 2nd order steady state solution}
For the forced solution we can use again the fact that the
statistics of the velocity field has no correlation with the forcing
of the passive scalar field at any time. Therefore we have:
\begin{equation}\label{forced1}
\langle |\theta_n(t)|^2\rangle_f=\int_0^{t}\sum_k {\cal P}^{(2)}_{n|k}(t-t')\langle
|f_k(t')|^2\rangle dt'\ .
\end{equation}
We should think of this equation only in the limit of $t\to \infty$,
since we need to eliminate the effects of exponentially decaying
initial value terms that do not contribute to the stationary forced
correlation function.
With a force which is Gaussian white noise, we write  $\langle
|f_k(t)|^2\rangle=f^2 \delta_{k,m}$. Using Eq. (\ref{Pmform}) for the propagator we get:
\begin{eqnarray}\label{forced2}
\langle |\theta_n(t)|^2\rangle_f\propto f^2&&\int_0^{t-\tau_m}H(n-\tilde m(t-t',m))
\\ &\times&2^{-\zeta_2(m-\tilde m(t-t',m))} dt' \nonumber \ .
\end{eqnarray}
We remind the reader that $\tau_m$ is the time it takes for the initial $\delta$-function to
develop a ``scaling tail'' for $n>m$, and now $m$ is the shell at which the random
forcing is localized.
The idea here is to use the fact that we know how to eliminate the time
variable in favor of the moving scale variable $\tilde m(t,m)$. Changing  variables of
integration to
$\tilde m$, using Eq. (\ref{mform}) we can write explicitly for $\tilde m\ll 
n$:
\begin{eqnarray}\label{forced3}
&&\langle |\theta_n(t)|^2\rangle_f\propto  \zeta_2 \ln(2)f^2\\&&\times\int_{-\infty}^{m}
2^{-\zeta_2(n-\tilde m)}2^{-\zeta_2(m-\tilde m)} 2^{-\zeta_2 \tilde m} d\tilde m
\nonumber
\ .
\end{eqnarray}
Note that we have extended in a formal manner the range of shell indices all the
way to $-\infty$, to allow for a long development of a self similar
solution. Naturally, since the integral converges quickly, this is
immaterial.  Finally, using (\ref{F2def}):
\begin{equation}\label{forced4}
F^{(2)}_n= {\rm Const}\times
2^{-\zeta_2 n} \, .
\end{equation}
This solution has the expected $2^{-\zeta_2 n}$, and is time
independent.

We note at this point that the forced solution $F^{(2)}_n$ had been shown
to be a {\em left} eigenfunction of eigenvalue 1 in Eq. (\ref{evalsum}).
Thus the first two subsections together fully demonstrate the two conjectures
(i) and (ii) from the introduction for the case of the second order
objects.
%%%%%%%%%%%%%%%%%%%%%%%%%%%%%%%%
\subsection{Why this simple time dependence?}

The knowledgeable reader might have noticed at this point that 
the explicit time dependence of the 2nd order propagator,
as displayed in Eq. (\ref{Pform}) is very simple. The exponent
of time, $t^{-1}$, is not anomalous, and appears independent of 
the second order exponent of the velocity field. This is not so
in the understood example of the Kraichnan model of passive
scalar advection, in which the time dependence of the operator
is anomalous \cite{RHK,98BGK}. To clarify this point we turn to the analysis
of the passive scalar shell model driven by a $\delta$-correlated
velocity field \cite{Luca}. In other words, for the velocity field $u$ in 
Eq. (\ref{passive}), we use a Gaussian field, $\delta$-correlated in time,
that satisfies:
\begin{eqnarray}\label{Vcort}
&&\langle u_n(t)u^*_m(t') \rangle=\delta_{n,m}\delta(t-t')C_n \, , \nonumber\\
&&C_n=C_0 2^{-\xi n}\, ,
\end{eqnarray} 
The calculations are described in Appendix \ref{Kraichnan}, with the
following results:
\begin{eqnarray}\label{Pext1}
\frac{d}{dt}\langle|\theta_{n}(t)|^2\rangle=M^{(2)}_{n,m}\langle|\theta_{m}(t)|^2\rangle
\end{eqnarray}
Where the matrix $\B.M^{(2)}$ is given by 
\begin{equation}\label{Mdef}
M^{(2)}_{n,m}=-\frac{2\delta_{n,m}}{\tau_n^{-1}+\tau_{n+1}^{-1}}
+\frac{2\delta_{m,n+1}}{\tau_n}+\frac{2\delta_{m,n-1}}{\tau_{n+1}} \ .
\end{equation}
Here $\tau_n\equiv 2^{-(2-\xi)n}/k_0C_0$. 
Since this matrix is time independent we have
\begin{equation}\label{Mexp}
{\cal P}^{(2)}_{n|m}(t)=[\exp(t M^{(2)})]_{n,m}\, .
\end{equation}
It is straigthforward to check that the 2nd order forced solution
$\langle |\theta_n|^2\rangle_f \sim 2^{-(2-\xi)n}\sim \tau_n$ is a zero mode
of $ \B.M^{(2)}$. In this case it is also straighforward to prove
that $I^{(2)}$ in Eq. (\ref{I2def}) is a conserved variable (in the infinite system limit). To do
so we note that on the one hand from Eq. (\ref{Pdt}) we have the following exact equation:
\begin{equation}\label{Pdisp}
\frac{d}{dt}\sum_{n=1}^{d}\langle|\theta_n(t)|^2\rangle
=-\frac{\langle|\theta_d(t)|^2\rangle}{\tau_{d+1}}\, .
\end{equation}

The explicit form of the quantity $I^{(2)}$ is in this case
\begin{equation}
I^{(2)} = \sum_m\tau_{m} \langle|\theta_m(t)|^2\rangle
\end{equation}
The rate of change of this object is
\begin{equation}\label{wei1}
\frac{d}{dt}\sum_m\tau_{m}
\langle|\theta_m(t)|^2\rangle=\sum_{m,n} \tau_mM^{(2)}_{m,n}
\langle|\theta_n(t)|^2\rangle
\end{equation}
Using the properties of $\B.M^{(2)}$ we can write:
\begin{equation}\label{weidisp}
\frac{d}{dt}\sum_n\tau_{n}\langle|\theta_n(t)|^2\rangle=-\langle|\theta_d(t)|^2\rangle\, .
\end{equation}

Taking the ratio of Eq. (\ref{Pdisp}) and  Eq. (\ref{weidisp}) we see that 
for the limit $d \rightarrow \infty$ the quantity $I^{(2)}$ is conserved with respect to 
the sum $\sum_m\langle|\theta_m(t)|^2\rangle$.

Now write the propagator in the form pertianing to $t\gg \tau_m$,
\begin{equation}
{\cal P}^{(2)}_{n|m}(t)=c\Big(\frac{\tau_m}{t}\Big)^\alpha H(n-\tilde m(t,m^*))\ , \label{trial}
\end{equation}
with 
\begin{equation}\label{mstar}
\tilde m(t)=m_0-\frac{1}{2-\xi}\log_2(t/\tau_m) \ .
\end{equation}
Write now the conservation law just proven as 
\begin{equation}\
I^{(2)} = \sum_n \tau_n {\cal P}^{(2)}_{n|m}(t) \approx {\rm Const} .
\end{equation}
Using the form (\ref{trial}) we require
\begin{equation}
\sum _n H(n-\tilde m(t,m^*))2^{-(2-\xi)(n-\alpha \tilde m(t,m^*))} ={\rm Const} \ .
\end{equation}
Obviously, this is constant iff $\alpha=1$, demonstrating the 
point that the explicit time dependence in our propagator is 
not anomalous. 
%%%%%%%%%%%%%%%%%%%%%%%%%%%%%%%%%
\section{The 4th order propagator}
\subsection{Simulations}
The 4th order propagator is defined by:
\begin{equation}
\langle |\theta_n(t)|^2|\theta_m(t)|^2\rangle=
{\cal P}^{(4)}_{n,m|p,q}(t)\langle |\theta_p(0)|^2|\theta_q(0)|^2\rangle \ .
\end{equation}
We remind the reader that the LHS has also contributions from other
initial conditions, i.e.  $\langle
\theta_{n+2}(0)\theta^*_{n+1}(0)\theta^*_{n+1}(0)\theta_{n-1}(0)\rangle$ but
these contributions appear in the numerics to be very small, and will not be
considered in this paper.   
For $\delta$-function initial conditions (say on  shell $p$) it is sufficient to consider
${\cal P}^{(4)}_{n,m|p,p}(t)$. For $m,n\ll p,q$ and for large times, ${\cal
P}^{(4)}_{n,m|p,q}(t)$ is indistinguishable from
${\cal P}^{(4)}_{n,m|p,p}(t)$.

First we studied the typical time dependence of the operator via
direct simulations. 
%%%%%%%%%%%%%%%%%%%%%%%%
%%%%%%%%%%%%%%%%%%%%
\begin{figure}
\hskip -0.5cm
\epsfxsize=8truecm
\epsfysize=8truecm
\epsfbox{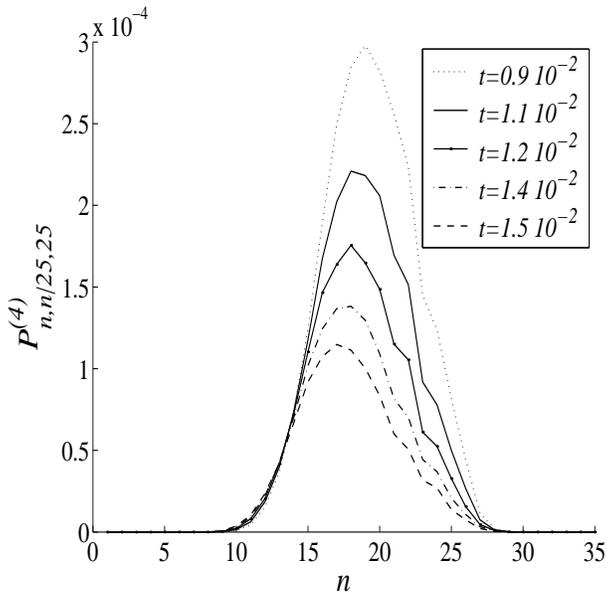}
\caption{The diagonal elements of ${\cal P}^{(4)}_{n,n|25,25}(t)$ as functions of
$n$ for five different times. The simulations were performed with
30 shells.}
\label{Pnnpp17}
\end{figure} 
%%%%%%%%%%%%%%%%%%%
%%%%%%%%%%%%%%%%%%%%%%%%
In Fig. \ref{Pnnpp17} we plot the diagonal elements ${\cal P}^{(4)}_{n,n|25,25}(t)$
as functions of $n$ for different times. Note the movement of the peak and
the decay of the function. This is very similar to what we found for
the second order propagator. In order to proceed we need to guess
an analytic form for the propagator and compare it with the numerical
data.

Our ansatz for the 4th order propagator is constructed using the
fusion rules \cite{96LP}. For the forced 4th order correlation 
functions the fusion rules predict that asymptoticaly for $|n-m|\gg 1$
\begin{equation}
F^{(4)}_{n,m}\propto 2^{-\zeta_4 \min(m,n)}2^{-\zeta_2|m-n|} \ . \label{F4fuse}
\end{equation}
This form was amply tested and demonstrated for shell models in \cite{99BBCT}.
It was shown that the asymptotic form is obtained very rapidly, for
any $|n-m| \ge 1$. Accordingly we expect that 
\begin{equation}\label{p4}
{\cal P}^{(4)}_{n,m|p,p}(t)=
\Big(\frac{\tau_p}{t}\Big)^{\zeta_4/\zeta_2}\!\!G\Big(\min(m,n)-\tilde
m(t,p)\Big)2^{-\zeta_2|m-n|}
\end{equation}
where the function $\tilde m(t,p)$ is the same as in Eq. (\ref{mform})
but with $p$ replacing $m$. 
The function $G(x)$ is expected to have, for $x\gg 0$, the scaling form:
\begin{eqnarray}\label{Gkform}
G(x)\propto 2^{-\zeta_4 x} \ .
\end{eqnarray}

\begin{figure}
\hskip -0.5cm
\epsfxsize=8truecm
\epsfysize=8truecm
\epsfbox{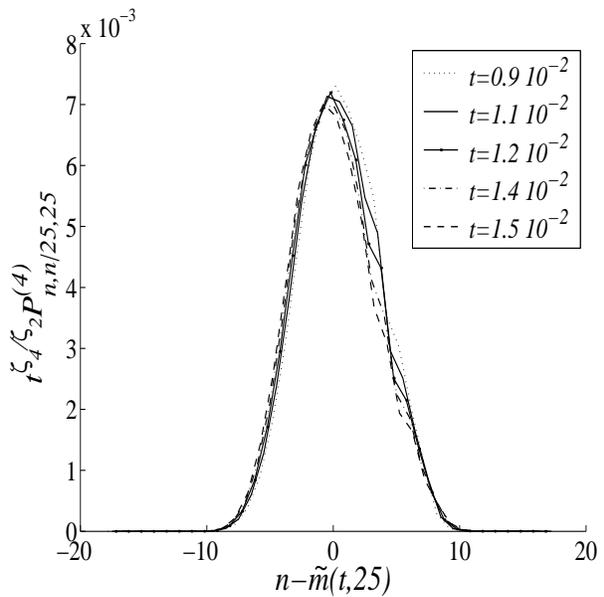}
\caption{The diagonal elements of $t^{\zeta_4/\zeta_2} {\cal P}^{(4)}_{n,n|25,25}(t)$
as a function of $n-\tilde m(t,25)$.}
\label{tweightPnnpp}
\end{figure}

The form (\ref{p4}) is very well supported by the data.
In Fig. \ref{tweightPnnpp} we replot the data of Fig. \ref{Pnnpp17}
multiplied by $t^{\zeta_4/\zeta_2}$, 
as a function of $n-\tilde m(t,25)$, where $\tilde m(t,25)$ solves
Eq. (\ref{mform}). It is obvious that the form (\ref{p4}) is justified
for the diagonal. 
%%%%%%%%%%%%%%%%%%%%
%%%%%%%%%%%%%%%%%%%  

\begin{figure}
\epsfxsize=6truecm
\epsfysize=6truecm
\epsfbox{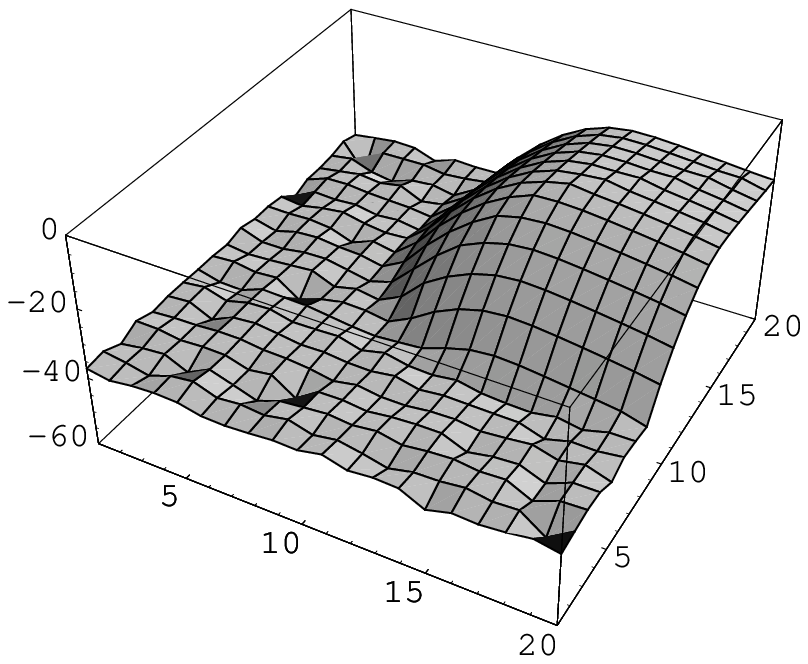}
\epsfxsize=6truecm
\epsfysize=6truecm
\epsfbox{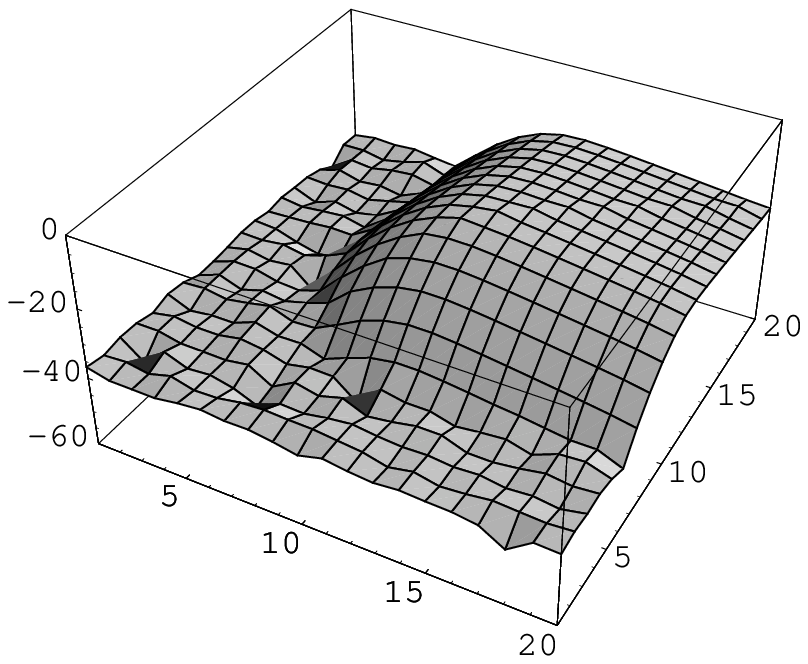}
\epsfxsize=6truecm
\epsfysize=6truecm
\epsfbox{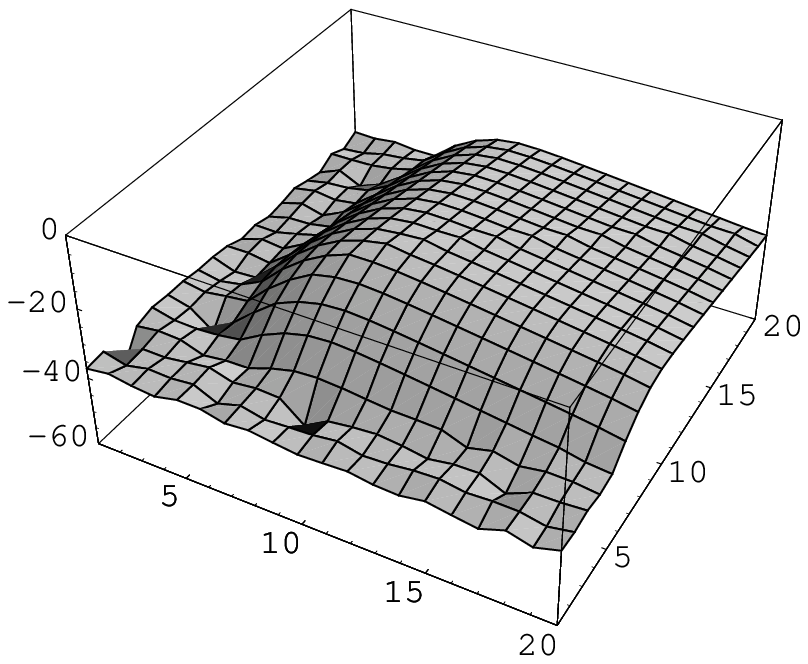}
\caption{The logarithm of the elements of $ {\cal P}^{(4)}_{n,m|18,18}(t)$ for the
Kraichnan shell model
as a function of $n$ and $m$ for the three different times $1.54\times 10^{-6}\tau_0$
(panel a),
$1.67\times 10^{-5}\tau_0$ (panel b) and $1.74\times 10^{-4}\tau_0$ (panel c).}
\label{Kraichdecay}
\end{figure}

\begin{figure}
\epsfxsize=6truecm
\epsfysize=6truecm
\epsfbox{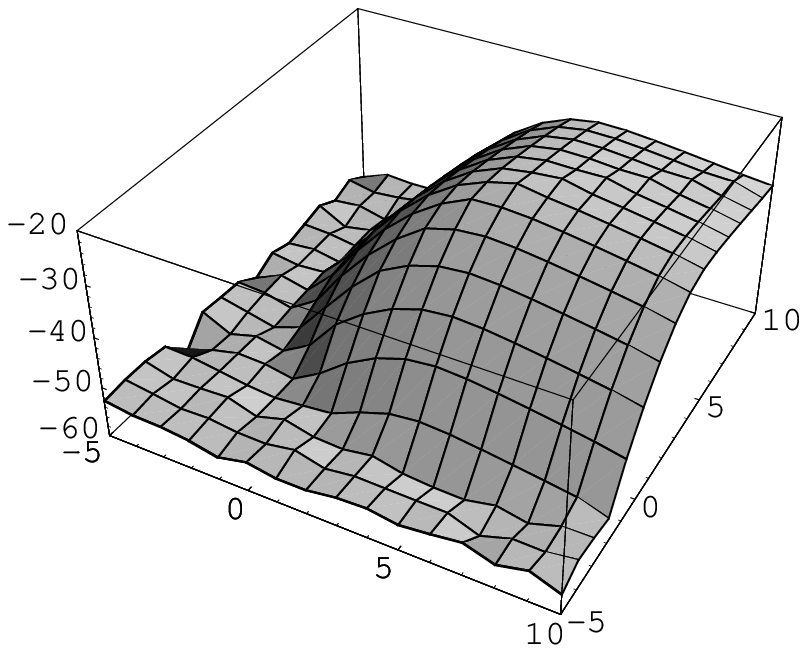}
\epsfxsize=6truecm
\epsfysize=6truecm
\epsfbox{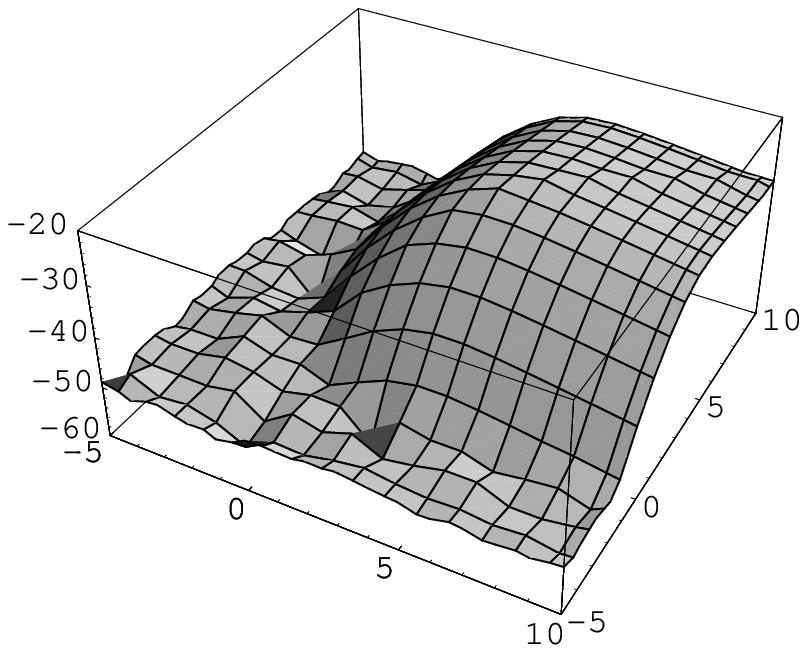}
\epsfxsize=6truecm
\epsfysize=6truecm
\epsfbox{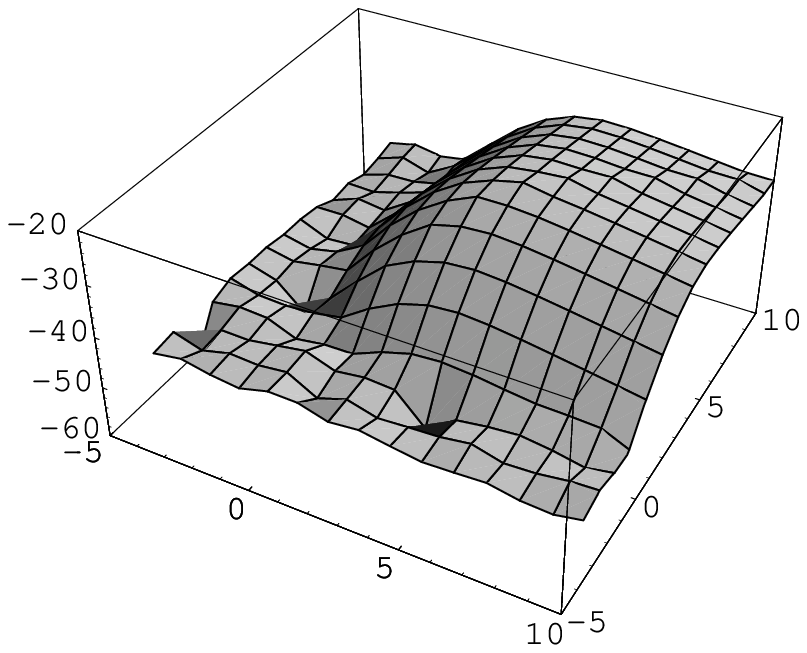}
\caption{The logarithm of the elements of $t^{\zeta_4/\zeta_2} {\cal P}^{(4)}_{n,m|18,18}(t)$ for the
Kraichnan shell model
as a function of $\big(n-\tilde m(t,18), m-\tilde m(t,18)\big)$. The
times are the same as in Fig. \ref{Kraichdecay}.
The invariance of the function is obvious.}
\label{Kraichtweight}
\end{figure}

It is more difficult to
demonstrate the full tensor by direct simulations; the off-diagonal elements are more noisy,
and the scaling behavior is somewhat less apparent than on the diagonal.
We can however obtain much better data for the Kraichnan model, for which 
${\cal P}^{(4)}_{n,m|p,p}(t)$ can be computed essentially analytically. In
Appendix \ref{Kraichnan} we present the derivation. Here we show
in Fig. \ref{Kraichdecay}  ${\cal P}^{(4)}_{n,m|18,18}(t)$ for three
different times. The spread and decay are apparent. In Fig. \ref{Kraichtweight}
the same data is shown after multiplying it by $t^{\zeta_4/\zeta_2}$,
and replotting it as a function of 
$\big(n-\tilde m(t,18), m-\tilde m(t,18)\big)$. Now the function is
preserved with respect to time.
%%%%%%%%%%%%%%%%%%%%
%%%%%%%%%%%%%%%%%%%  
%%%%%%%%%%%%%%%%%%%%
%%%%%%%%%%%%%%%%%%%  
 
%%%%%%%%%%%%%%%%%%%%%%%%%%%%
\section{Consequences of the form of the 4th order propagator}

\subsection{The 4th order constant of the motion}

\begin{figure}
\epsfxsize=8truecm
\epsfbox{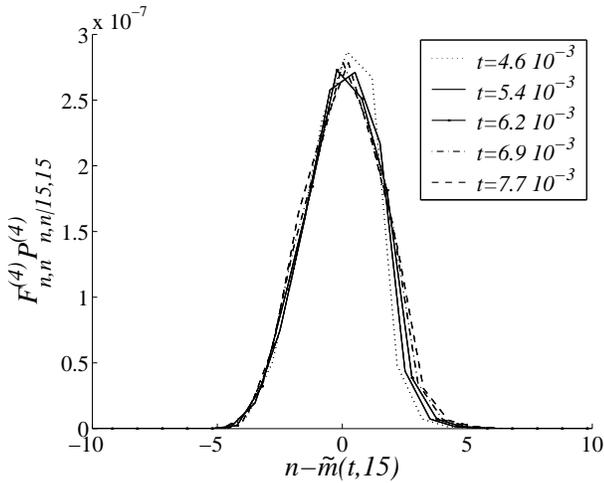}
\caption{The weighted elements $F^{(4)}_{n,n}{\cal P}^{(4)}_{n,n|15,15}(t)$
as a function of $n-\tilde m(t,15)$.}
\label{new}
\end{figure}

According to the conjectures discussed in the introduction
(in particular Eq. (\ref{conj2})),
we expect the forced solution $F^{(4)}$ to act as
the {\em left} eigenfunction of eigenvalue 1, $Z^{4}$.
Here we demonstrate that $I^{(4)}$ as defined by
Eq. (\ref{I4def}) is indeed a constant of the motion.
Using for $F^{(4)}$ Eq. (\ref{F4fuse}), and
expressing $t^{\zeta_4/\zeta_2}$ in terms of $\tilde m$ we get:
\begin{eqnarray}
I^{(4)}(t)&=&\sum_{n,m}
G\Big(\min(m,n)-\tilde m(p,t)\Big) \nonumber \\ &\times&
2^{-2\zeta_2|m-n|}2^{-\zeta_4(\min(m,n)-\tilde m(p,t))}
\end{eqnarray}
As in Eq. (\ref{evalsum}), the time dependance of the sum is eliminated
because time is introduced only through the expression $\min(m,n)-\tilde
m(p,t)$. Consequently the object $I^{(4)}$ becomes time independent.
We demonstrate this invariance for the diagonal part of the summand
in Fig. \ref{new}. 
%%%%%%%%%%%%%%%%%%%%
%%%%%%%%%%%%%%%%%%%   
To display the invariance for the whole weighted tensor 
we employ again the data presented in
Fig. \ref{Kraichdecay}. After multiplication by the weights $F_{n,m}^{(4)}$
and replotting in moving coordinates, the constancy of the summand of $I^{(4)}$ is
demonstrated. This is done in Fig. \ref{KraichF4}, using the analytic results of Appendix
\ref{Kraichnan}.
%%%%%%%%%%%%%%%%%%%%
\begin{figure}
\epsfxsize=6truecm
\epsfysize=6truecm
\epsfbox{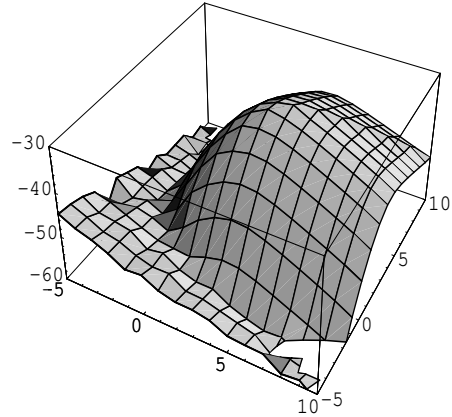}
\epsfxsize=6truecm
\epsfysize=6truecm
\epsfbox{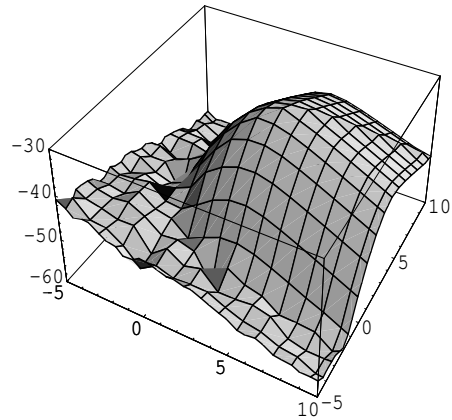}
\epsfxsize=6truecm
\epsfysize=6truecm
\epsfbox{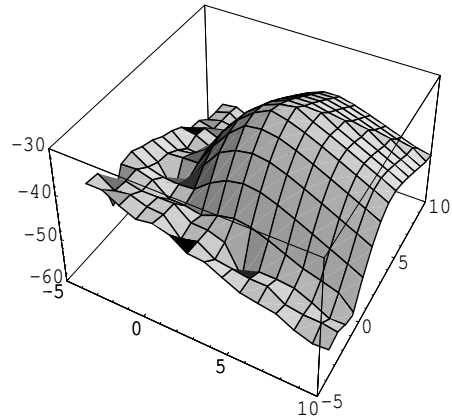}
\caption{The elements of $F^{(4)}_{n,m} {\cal P}^{(4)}_{n,m|18,18}(t)$ for the
Kraichnan shell model
as a function of $\big(n-\tilde m(t,18), m-\tilde m(t,18)\big)$. The
times are the same as in Fig. \ref{Kraichdecay}.
The invariance of the function is obvious.}
\label{KraichF4}
\end{figure} 
%%%%%%%%%%%%%%%%%%%  
 
%%%%%%%%%%%%%%%%%%%
\subsection{The forced 4th order steady state solution}
Finally, we can calculate the analog of Eq. (\ref{forced1}), for the steady
state 4-point function in a system forced by gaussian white noise. 
Returning to Eq. (\ref{Rnm}) we write
\FL
\begin{eqnarray}
&&F^{(4)}_{n,m}=\int_0^t\cdots \int_0^t
ds_1\cdots ds_4 \langle R_{n,p}(t|s_1) R^*_{n,p'}(t|s_2)\nonumber\\
&&R_{m,q}(t|s_3)
R^*_{m,q'}(t|s_4)\rangle
\langle f_p(s_1)f^*_{p'}(s_2)f_q(s_3)f^*_{q'}(s_4)\rangle  \ , \label{forced}
\end{eqnarray}
where we have used the statistical independence of the forcing from
the velocity field. We note that in Eq. (\ref{forced}) the
time integration can be (and should be) extended to arbitrarily long times, to get
a stationary forced correlation function. This way we also get rid of exponentially 
decaying initial value terms. Using the correlation properties of the forcing Eq. (\ref{ff})
we obtain 
\begin{equation}
F^{(4)}_{n,m}(t)=2 C_p C_q\int_0^t\!\!\!\!\! ds_1 \int_0^t
\!\!\!\!\!ds_2 \langle |R_{n,p}(t|s_1)|^2 |R_{m,q}(t|s_2)|^2 \rangle \ .
\end{equation}
We next split the integral into two domains in which $s_2\le s_1$ and 
the opposite. Consider the first domain in which the integral on
the RHS has the form
\begin{eqnarray}
&&\int_0^t\!\!\!\!\! ds_1 \int_0^{s_1}
\!\!\!\!\!ds_2 \langle |R_{n,p}(t|s_1)|^2 |R_{m,q}(t|s_2)|^2 \rangle \nonumber\\
= &&\int_0^t\!\!\!\!\! ds_1 \int_0^{s_1}
\!\!\!\!\!ds_2 \langle |R_{n,p}(t|s_1)|^2 |R_{m,\ell}(t|s_1)|
^2|R_{\ell,q}(s_1|s_2)|^2\rangle\ .
\end{eqnarray}
To proceed we need to consider the decay time of the operator
$R_{n,m}(t|t_0)$ compared to the decorrelation properties of
products of such operators {\em at different times}. On the one hand we know that
these operators depend explicitly on time, decaying like a power of time
(cf. Eq. (\ref{p4})). On the other
hand, we  expect the correlation of products of
different time operators to decay exponentially, since the
operators $R_{n,m}(t|t_0)$ contain the chaotic velocity field that
appears in the exponential, cf. Eq. (\ref{Rnm}). The time domain
is arbitrarily long, but throughout most of the time integration
the product is actually decorrelated, and we can write
\begin{eqnarray}
&&\int_0^t\!\!\!\!\! ds_1 \int_0^{s_1}
\!\!\!\!\!ds_2 \langle |R_{n,p}(t|s_1)|^2 |R_{m,\ell}(t|s_1)|
^2|R_{\ell,q}(s_1|s_2)|^2\rangle\nonumber\\
&&\approx\int_0^t\!\!\!\!\! ds_1 \langle |R_{n,p}(t|s_1)|^2 |R_{m,\ell}(t|s_1)|
^2\rangle \int_0^{s_1}
\!\!\!\!\!ds_2 \langle |R_{\ell,q}(s_1|s_2)|^2\rangle \ .
\end{eqnarray}
We can now perform the integral over $s_2$, with a result
independent of either $s_1$ or $t$. Finally, if we choose the forcing in 
Eq. (\ref{forced}) as a single shell
forcing on the shell $p$, $\langle |f_k(t)|^2 |f_\ell(t)|^2\rangle
=C_p^4\delta_{k,p}\delta_{\ell,p}$ we get:
\begin{eqnarray}
F^{(4)}_{n,m}&\propto&2^{-\zeta_2|m-n|}
\int_0^{t-\tau_p}\Big(\frac{\tau_p}{t-t'}\Big)^{\zeta_4/\zeta_2}
\nonumber\\ &\times&2^{-\zeta_4(\min(m,n)-\tilde m(t-t',p))}dt' \ ,
\end{eqnarray}
where we have used the analytic asymptotic form of ${\cal P}^{(4)}_{n,m|p,p}(t)$,
Eq. (\ref{p4}).
Changing the integration variable to $\tilde m(t-t',p)$ we get:
\begin{eqnarray}
F^{(4)}_{n,m}&&\propto 2^{-\zeta_2|m-n|}2^{-\zeta_4(\min(m,n))}\nonumber \\
 &&\times\int_{-\infty}^{p}2^{2\zeta_4 \tilde m} 2^{-\zeta_2\tilde m}d\tilde m \ ,
\end{eqnarray}
or, we find a time independent solution:
\begin{equation}
F^{(4)}_{n,m}={\rm Const} \times 2^{-\zeta_2|m-n|}2^{-\zeta_4\min(m,n)}
\end{equation}
As expected, this is the correct form of the 4th order correlation function, in agreement
with the fusion rules. 

The theory for the sixth and higher order correlation functions follows
the same lines, ane will not be reproduced here.
%%%%%%%%%%%%%%%%%%%%%%%%%%%%
\section{Summary and Concluding Remarks}

In summary, we examined in detail the statistical physics of the
shell model of a passive scalar advected by a turbulent velocity
field. We presented a theory to explain and solidify the two conjectures
proposed in \cite{01ABCPV} and reproduced in the introduction. These
conjectures state that (i) in the decaying problem there exist infinitely many statistically
conserved quantities, denoted above as $I^{(N)}$; (ii) These quantities are obtained
by  integrating (or summing) the decaying correlation functions against
the stationary correlation functions of the forced problem. We have pointed
out that the conjectures imply that the forced solutions are 
{\em left}-eigenvectors of eigenvalue 1 of the propagators $\B.{\cal P}^{(N)}$.
For the model discussed above we have established these conjectures
by examining the form of the propagators $\B.{\cal P}^{(N)}$. Using numerical
simulations as a clue, we proposed analytic expressions for the 
operators $\B.{\cal P}^{(2)}$ and $\B.{\cal P}^{(4)}$, pointing out that
similar concepts (fusion rules \cite{96LP} in particular) can be used to write
down also the higher order operators. We checked the analytic forms
against the simulations, and proceeded to demonstrate that the
forced, stationary correlation functions are indeed left-eigenvectors
of eigenvalue 1 of these operators. This implies that the objects
$I^{(N)}$ are indeed constants of the motion. Next we derived
the forced, stationary correlation functions, and showed that
the form of our operators dictates scaling solutions in agreement
with the fusion rules. As a result the two conjectures were confirmed.
In our analytic calculations we used repeatedly the fact that the
operators ``compactify" in shell-space once expressed in terms of
a single moving scale whose dynamics was determined analytically. 

One should state a caveat at this point: the analytic form of the 
opeartors $\B.{\cal P}^{(2)}$ and $\B.{\cal P}^{(4)}$ was {\em guessed}
on the basis of numerics and the fusion rules. Although they appear
to agree with the simulations, we cannot state that the forms are
{\em exact}. Accordingly, until these forms are derived from first
principles, the exact status of the conjectures is not established.
It may be that the conjectures are only satisfied to a good
approximation. This question needs to be addressed in future research.

Notwithstanding this caveat, we should point out a surprising bonus
of the approach discussed in this paper: we have at hand an analytic 
form of the propagators. {\em We can thus provide analytic
predictions for the decaying correlation functions for arbitrary
initial conditions}. Considering that the velocity field
is a solution of a highly non-trivial chaotic dynamical
system, and that the passive scalar is slaved to it, it is quite 
gratifying that nevertheless one can offer analytic solutions
for the time-dependent correlation functions of the latter.
It is of course very tempting to hope that a similar
theory can be developed in other cases of turbulent transport,
leading to analytic predictability of the time-dependent
correlation functions in the decaying case. Since this paper
demonstrated that the Lagrangian structure is not a prerequisite
for the exisitence of Statistically Preserved Structures, we feel
that such a theory should be sought in the Eulerian frame in which
calculations are much easier than in the Lagrangian frame. This
development should be addressed by future research.
%%%%%%%%%%%%%%
\acknowledgments
It is a pleasure to acknowledge extensive discussions with 
I. Arad, L. Biferale, A. Celani, K. Gawedzki, P. Muratore-Ginanneschi and
M. Vergassola. Special thanks are due to L. Biferale for sharing
with us his numerical results, and to L. Biferale, A. Celani and M. Vergassola for
a critical reading of a draft of this paper. This work has been
supported in part by the European Union under the TMR program
``nonideal turbulence", the German Israeli Foundation and the Naftali and Anna 
Backenroth-Bronicki Fund for Research in Chaos and Complexity.
TG thanks the Israeli Council for Higher Education and the Feinberg 
postdoctoral Fellowships program at the WIS for financial support.
%%%%%%%%%%%%%%%%%%%%%%%%%%%%%%%%
\appendix
\section{The Kraichnan shell model}
\label{Kraichnan}
For the velocity field $u$ in Eq. (\ref{passive}) we use a Gaussian, 
delta-correlated in time
field that satisfies:
\begin{eqnarray}\label{Vcor}
&&\langle u_n(t)u^*_m(t') \rangle=\delta_{n,m}\delta(t-t')C_n \, , \nonumber\\
&&C_n=C_0 2^{-\xi n}\, ,
\end{eqnarray} 
For this simple model we can find a closed form
equation for the time derivative of $\B.{\cal P}^{(2)}(t)$.
For simplicity we set the diffusivity $\kappa=0$, and
replace its effect by truncating the opeartor at the dissipative shell $d$ 
(cf. Eq. (\ref{defkd})).
%%%%%%%%%%%%%%%%%%%%%%%%%%%%%%%%%%%%
\subsection{The 2nd order operator}
We evaluate the 2nd order propagator's time derivative by
multiplying Eq. (\ref{passive}) by $\theta_n^*$ and adding the complex conjugate to get:
\begin{eqnarray}
&&\frac{d}{dt}\langle|\theta_n(t)|^2\rangle= i
k_{n+1}\langle u_{n+1}(t)\theta_{n+1}(t)\theta^*_n(t)\rangle\nonumber\\&&+i k_n \langle
u^*_{n}(t)\theta^*_{n}(t)\theta_{n-1}(t)\rangle+C.C \ , \label{Gaus3}
\end{eqnarray}
Using Gaussian integration by parts we compute the 3rd
order correlation functions including the velocity:
\begin{eqnarray}\label{Gaus1}
&&\langle\theta^*_n(t)\theta_m(t) u_m(t)\rangle=
\int dt'\sum_p \langle u_m(t)u^*_p(t')\rangle
\nonumber \\ &&\times \Big[\langle\frac{\delta \theta^*_n(t)}{\delta
  u^*_p(t')}\theta_m(t)\rangle+\langle\theta^*_n(t)\frac{\delta
  \theta_m(t)}{\delta u^*_p(t')}\rangle\Big]\, .
\end{eqnarray}
From Eq. (\ref{passive}) we have for the functional derivatives:
\begin{eqnarray}\label{Gaus21}
&&\frac{\delta \theta_p(t)}{\delta u^*_q(t')}=
i\Theta(t-t')\delta_{p,q}k_{q}\theta_{q-1}(t') \, ,\nonumber\\
&&\frac{\delta \theta^*_p(t)}{\delta u^*_q(t')}=
-i\Theta(t-t')\delta_{p+1,q}k_{q}\theta^*_{q}(t')\, ,
\end{eqnarray}
where $\Theta(t)$ is the step function, $\Theta(t) = 0$, $t<0$,
$\Theta(t)=1$,
$t>0$, $\Theta(0)=1/2$. Plugging Eqs.(\ref{Vcor}), (\ref{Gaus21}) into Eq. (\ref{Gaus1}), 
Eq. (\ref{Gaus3}) becomes
\begin{eqnarray}
\frac{d}{2dt}\langle|\theta_n(t)|^2\rangle&=&
C_{n+1}k^2_{n+1}\langle|\theta_{n+1}(t)|^2\rangle+
C_{n}k^2_{n}\langle|\theta_{n-1}(t)|^2\rangle\nonumber\\&-&
(C_{n}k^2_{n}+C_{n+1}k^2_{n+1})\langle|\theta_{n}(t)|^2\rangle
\ . \label{Pdt}
\end{eqnarray}
This can be written in matrix form as:
\begin{eqnarray}\label{Pext}
\frac{d}{dt}\langle|\theta_{n}(t)|^2\rangle=M^{(2)}_{n,m}\langle|\theta_{m}(t)|^2\rangle
\end{eqnarray}
Where the matrix $\B.M^{(2)}$ is given by Eq. (\ref{Mdef}).
It is time independent and thus a solution for
${\cal P}^{(2)}_{n|m}(t)$, defined in Eq. (\ref{forced1}), can be written as
Eq. (\ref{Mexp})
%%%%%%%%%%%%%%%%%%%%%%%%%%%%%%%%%%%%%%%%
\subsection{The 4th order operator}

Let us consider the propagator of the 4-point correlation function 
$\langle|\theta_n(t)|^2|\theta_m(t)|^2\rangle$~:
\begin{equation}
\frac{d}{dt}\langle|\theta_n(t)|^2|\theta_m(t)|^2\rangle
=M^{(4)}_{n,m,p,q}\langle|\theta_p(t)|^2|\theta_q(t)|^2\rangle\ ,
\label{dtM4}
\end{equation}
where the operator $\B.M^{(4)}$ can be computed in analogy to
Eqs. (\ref{Pdt}), (\ref{Pext}), (\ref{Mdef}):
\begin{eqnarray}
&&M^{(4)}_{n,m,p,q}=\nonumber\\&&\frac{1}{2}\Big(M^{(2)}_{n,p}\delta_{m,q}
+M^{(2)}_{n,q}\delta_{m,p}
+\delta_{n,p} M^{(2)}_{m,q}
+\delta_{n,q} M^{(2)}_{m,p}\Big)\nonumber \\
&&+2\tau^{-1}_n\Big(\delta_{n,m}-\delta_{n,m+1}\Big)
\Big(\delta_{n,p}\delta_{n-1,q}+\delta_{n,q}\delta_{n-1,p}\Big)\nonumber \\
&&+2\tau^{-1}_{n+1}\Big(\delta_{n,m}-\delta_{n+1,m}\Big)
\Big(\delta_{n,p}\delta_{n+1,q}+\delta_{n,q}\delta_{n+1,p}\Big) \ .
\label{evoleqF4}
\end{eqnarray}
We note that $\B.M^{(4)}$ is not symmetric under the exchange of left and right
indices, i.~e. $M^{(4)}_{n,m,p,q}\neq M^{(4)}_{p,q,n,m}$, and thus admits 
different left and right eigenvectors.
The zero-mode of Eq. (\ref{dtM4}) satisfies
\begin{equation}
M^{(4)}_{n,m,p,q}Y^{(4)}_{p,q} = 0\ ,\label{revM4}
\end{equation}
and is expected to be a symmetric function of the form
\begin{equation}
Y^{(4)}_{n,m}=2^{-\zeta_4\min(m,n)}f^{\rm R}(|m-n|)\ .\label{formrevM4}
\end{equation}
Equivalently one can
consider a {\em left} zero-mode of $\B.M^{(4)}$, which we denoted above by 
$\B.Z^{(4)}$ (cf. Eq. (\ref{eig})),
\begin{eqnarray}
&&M^{(4)}_{p,q,n,m}Z^{(4)}_{p,q} = 0\ ,\label{levM4}\\
&&Z^{(4)}_{n,m}=2^{-\zeta_4\min(m,n)}f^{\rm L}(|m-n|)\ .\label{formlevM4}
\end{eqnarray}
We will show that both left and right zero modes have overall scaling exponent $\zeta_4$,
multiplied by a function $f^{\rm R/L}(|m-n|)$,  which scales like
$2^{-\zeta_2|m-n|}$  provided $|m-n|\gg 1$, in agreement with the fusion rules. We therefore 
propose the following ansatz
\begin{equation}
f^{\rm R/L}(q)=
\sum_{j=1}^{\infty}a^{\rm R/L}_j\tau_q^{j}\ ,\quad\quad q>0\ .
\label{fexp}
\end{equation}
Plugging this ansatz into Eqs. (\ref{revM4}), (\ref{levM4}), we find three
different cases~: (i) $m=n$, (ii) $m=n \pm 1$ 
and (iii) $|m-n|>1$. This last case, which is identical for both left and
right equations, reads (assuming $m>n+1$)
\begin{eqnarray}
&&(\tau^{-1}_n+\tau^{-1}_m+\tau^{-1}_{n+1}+\tau^{-1}_{m+1})f^{\rm R/L}(m-n)
\nonumber\\
&&=\tau^{-1}_{m+1}f^{\rm R/L}(m-n+1)+\tau^{-1}_{m}f^{\rm R/L}(m-n-1)
\nonumber\\&&+\tau^{-1}_{n+1}2^{-\zeta_4}f^{\rm R/L}(m-n-1)+\tau^{-1}_{n}2^{\zeta_4}
f^{\rm R/L}(m-n+1)\ ,
\end{eqnarray}
which, defining $\beta=\zeta_4-2 \zeta_2$, yields the following recursion 
relation for the coefficients $a^{\rm R/L}_j$ in Eq. (\ref{fexp}),
\begin{equation}
a^{\rm R/L}_{j}=
-\frac{1+\tau^{-1}_1-2^{-\beta}\tau^{-1}_{j-2}-2^{\beta}\tau_{j-3}}
{1+\tau^{-1}_1-\tau_{j-1}-\tau^{-1}_{j}}a^{\rm R/L}_{j-1} \quad j\geq 2\ .
\end{equation}
It then remains to determine $\beta$, $f^{\rm R/L}(0)$ and $a^{\rm R/L}_1$,
which is done with the help of cases (i) and (ii) above. In the case of the 
right zero-mode, we have
\begin{eqnarray}
(1+\tau^{-1}_{1})f^{\rm
  R}(0)&=&2\tau^{-1}_{1}(1+2^\beta\tau^{-1}_{1})f^{\rm R}(1)\ ,\\
(1+4\tau^{-1}_{1}+\tau^{-1}_{2})f^{\rm R}(1)&=&
(\tau^{-1}_{1}+2^{-\beta}\tau_{1})f^{\rm R}(0)\nonumber\\
&&+(1+2^\beta)\tau^{-1}_{2}f^{\rm R}(2)\ ,
\end{eqnarray}
whereas the left zero-mode yields
\begin{eqnarray}
(1+\tau^{-1}_{1})f^{\rm L}(0)&=&\tau^{-1}_{1}(1+2^\beta\tau^{-1}_{1}) 
f^{\rm L}(1)\ ,\\
(1+4\tau^{-1}_{1}+\tau^{-1}_{2})f^{\rm L}(1)&=&
2 (\tau^{-1}_{1}+2^{-\beta}\tau_{1})f^{\rm L}(0)\nonumber\\
&&+(1+2^\beta)\tau^{-1}_{2}f^{\rm L}(2)\ .
\end{eqnarray}
We note that,  provided we impose $f^{\rm L}(0)=1/2 f^{\rm R}(0)$, which
amounts to fixing the arbitrary relative multiplicative factor between
$\B.Y^{(4)}$ and $\B.Z^{(4)}$, we obtain two identical solutions,
i.~e. $a^{\rm R}_j=a^{\rm L}_j$ $\forall j\ge 1$. The anomaly $\beta$ is
then the same for both systems of equations and can be determined numerically.
%%%%%%%%%%%%%%%%%%%%%%%%%%%%%%%%%%%%%%%%%%%%%%%%%%%%%%%%%%%%%%%%%%%%%%%%%%%%%%
%
% Bibliography
%
%%%%%%%%%%%%%%%%%%%%%%%%%%%%%%%%%%%%%%%%%%%%%%%%%%%%%%%%%%%%%%%%%%%%%%%%%%%%%%

\end{document}